\documentclass{article}

\usepackage{amsmath}
\usepackage{algorithm}
\usepackage[noend]{algpseudocode}

\usepackage[indent=15pt, parfill]{parskip}
\usepackage{indentfirst}
\title{An O($nlogn$) approximate knapsack algorithm}
\author{Nick Dawes, Ottawa, Canada: nwdawes314@gmail.com}
\date{December 23, 2025}
\begin{document}

\maketitle

\section {Abstract}

A modified dynamic programming algorithm rapidly and accurately solves large 0/1 knapsack problems. It
has computational O($nlogn$), space O($nlogn$) and predictable maximum error.
Experimentally it's accuracy increases faster than linearly with the solution size $k$ 
(the number of objects placed in the knapsack).
Problems with $k=10^3$ are solved with an average maximum fractional error of $10^{-4}$ 
and problems with $k=10^5$ with an average maximum fractional error of $10^{-7}$.
The algorithm runs in constant time for all problems with a given $n$.  
On a common desktop computer the algorithm processes $n=10^3$ problems in $10^{-3}$ seconds 
and $n=10^6$ problems in 2 seconds.

\section {Introduction}

Given a multiset of $n$ objects with profits $p_i$ and weights $w_i$ , the knapsack problem is to select the subset with maximum profit and whose weight $<= c$ (the knapsack weight capacity). The $0/1$ form of this problem is where each object can only be used once.  It has many applications in resource allocation, logistics and finance. The real number form of this problem is NP-complete [KA72].There are two classes of algorithms which solve it: exact and approximate. We are mostly concerned with approximate algorithms in this work. 

Many authors have described, analysed and discussed knapsack problems; for example Dantzig [DA56], Schroeppel and Shamir[SS81] and Pisinger [PI99]. 
Bellman's classic dynamic programming [BE53] was applied in pseudopolynomial time O($nT$). 
Branch and bound methods have been widely applied, significantly by Martello et al [MP99].
Knapsack approximation algorithms were described and analysed early on by Sahni [SS75], Ibarra and Kim [IK75] and since then by many others [CL03], [KK04],  [BL15], [HK17], [MM22], [KA23].
FPTAS (fully polynomial time approximation schemes) have been found, recently with improved performance to O($n+(1/e)^2$) by Chen et al [CL25] ($e$ is the fractional error in the optimal subset profit).
The maximum profit possible for a problem instance has been defined by Pisinger [PI99]. 
Sahni [SS75] analysed approximate algorithm performance for small n problems.
An analysis of theoretical errors for various forms of greedy algorithms was reported by Calvin and Leung[CL03].
They showed under some conditions the greedy algorithm's accuracy should increase at $n^{-1/2}$. 
Pisinger [PI99] and Jooken [JO22] described hard problem generators. Jooken also provided hard problem instances with their optimal solutions.

\subsection {Trials and tests}

Two different types of knapsack problem instance testing were performed: "random trials" and "hard file tests".
Random trials consisted of randomly generated real $p_i$ and $w_i$ in the range $0..1$ for a selected $n$. The capacity $c$ was chosen as a random fraction of the sum of the weights. The capacity upper limit was 90\% of the sum of the weights. Its lowest limit was the weight of the lightest object.  For "fixed $k$ random trials",
$k$ was fixed for each random trial instance by varying $c$ until the optimal subset size found by greedy was within 1\% of that required. 
Hard file tests were the 2977 problem instance files from Jooken [JO22] for which optimal solutions had been recorded.

All the logic given below was implemented in C++. 
All object values, knapsack capacities and all real number operations were carried out in double precision.
The 64 bit Mersenne twister prng was used. 
All experiments ran on one core of an 8Gb 4-core 3.5ghz Windows 10 machine, with no parallel logic.

\section {XDP}

The well known dynamic programming algorithm for knapsack with $T$ bins and O($nT$) time complexity has been significantly changed. Five changes let XDP run far faster and in $O(nlogn)$ space to achieve similar accuracy.  The pseudo code for XDP is given below in Algorithm 1. 

The first change is that $T=ln(n) * g$, where $g$ is a constant. Therefore the computational complexity for $n$ objects processing $T$ bins is $O(nlogn)$.  The constant $g$ was selected $g=12$ so that the runtime for this algorithm is the same as the $O(nlogn)$ initial sort of objects by $p_i/w_i$.

Each bin records data for the subset with the largest profit whose sum of weights maps onto that bin. The objects are presorted by $p_i/w_i$, high to low. Each object is processed against the data for each bin with a subset. The second change is that bin indexing does not use scaled and rounded $w_i$ data, it uses the exact weight of the subset. Accordingly the bin index = $subset weight * (T/c)$.

The third change is the addition of backtracking data and logic to extract the optimal subset's members. This data and logic is perhaps most easily understood from the pseudo code. An array $back[i][j]$ is the previous object index added to the subset to which object $i$ is now added to produce a new best subset in bin $j$. Backtracking starts from data for the optimal subset and steps backwards using $back[i][j]$ until it reaches the start of this subset. $back[i][j]$ has space $O(nlogn)$.

The fourth change is that all objects are processed, even the lightest.

The fifth change is the calculation of the maximum error. The maximum possible sum of profits $Pmax$ for this problem instance is computed by a modified greedy algorithm, whose details are in Appendix 1. For an XDP optimal profit of $S$ on this instance, XDP's maximum error $e =(Pmax - S)/Pmax$.

\begin{algorithm}
\caption{XDP:}

\begin{algorithmic}[1]

\Procedure{XDP}{}

\State $T \gets ln(n) * g$ \Comment{T is the number of bins.  g=12}
\State $bestbin \gets0$ \Comment{The bin with the best subset}
\State $S \gets 0$ \Comment{best sum of profits of any subset}
\State $XO[1..T] \gets-1$ \Comment{these bins are empty}
\State $XO[0] \gets 0$ \Comment{let objects start a subset from bin 0}
\State $XP[0..T] \gets 0$ \Comment{sum of the profits for the best subset in this bin}
\State $XW[0..T] \gets 0$ \Comment{sum of the weights for the best subset in this bin}
\\

\For {$ i=1,n$}
\For {$ j=T,0$}	\Comment{process bins high to low}
\If {$XO[j] >= 0$}  \Comment{this bin contains data for a subset}
\State $a \gets XP[j] + p_i$  \Comment{the proposed subset's profit sum}
\State $b \gets XW[j] + w_i$ \Comment{the proposed subset's weight sum}
\\
\If {$b<= c$}  	\Comment{the knapsack can hold this subset's weight}
\State $k \gets b * T/c$  \Comment{k is the target bin: note: k can equal j}

\\
\If {$a > XP[k]$}   \Comment{a new best subset for bin $k$}
\State $XP[k] \gets a$
\State $XW[k] \gets b$
\State $back[i][k] \gets XO[j]$
\State $XO[k] \gets i$ 
\If {$a > S$} 
\State $S \gets a$
\State $bestbin\gets k$
\EndIf
\EndIf
\EndIf
\EndIf
\EndFor
\EndFor
\\
\State $e \gets (Pmax - S)/Pmax$ \Comment {maximum error using $Pmax$ from Greedy}
\\
\\
\textit{Back track to determine the best subset}
\\
\State $x[1..n]\gets 0$	\Comment{no objects are in the best subset}
\State $b \gets XW[bestbin]$ \Comment{the best subset's weight sum}
\State $i \gets XO[bestbin]$\Comment{the object which defined the best subset's profit}
\State $k \gets bestbin$ 
\While {$i>0$}
\State $x]i] \gets 1$ \Comment{add object $i$ to the best subset}
\State $b \gets b-w_i$ 	\Comment{the source subset's weight sum}
\State $i \gets back[i][k]$
\State $k \gets b * T/c$	
\EndWhile
\EndProcedure
\end{algorithmic}
\end{algorithm}

\subsection {XDP: results and discussion}

Table 1 reports the average maximum fractional error $e$ of XDP with fixed $k=50$, averaged over 1000 fixed k random trials for each $n=100..10^5$. This table shows $e$ from XDP is nearly independant of $n$ for $k=50$. 

Table 2 reports the average maximum fractional error $e$ of XDP with fixed $k=5..50,000$, averaged over $n=10..10^5$. 1000 fixed k random trials were carried on each possible $n=10, 10^2, 10^3, 10^4$ and $10^5$. Also shown is $e.k$. If $e$ varied only inversely with $k$, this column would be a constant. Since $e.k$ drops as $k$ rises, $e$ from XDP drops faster than linearly with $k$.

Table 3 reports the average maximum fractional error $e$ and runtime of XDP, averaged over 1000 random trials for each $n=10..10^6$. 
The average $k$ for these trials is also shown for each $n$.
$e$ for each $n$ in Table 3 can be compared to the entry with the nearest corresponding $k$ in Table 2. For example, in Table 3 the average $k$ for $n=10^3$ is $6.0* 10^2$. The nearest entry to $k= 6.0* 10^2$ in Table 2 is the entry for $k= 5* 10^2$, which has $e=1.13 * 10^{-4}$. $e=1.58 * 10^{-4}$ from Table 3's entry is very similar.
Such correspondance of Tables 2 and 3 shows $e$ from XDP depends on $k$ and is nearly independant of $n$. This dependance of $e$ on $k$ rather than $n$ is further discussed in Appendix 2.

Table 3 also shows XDP is practical and accurate on very large problems. The time to process $n=10^6$ problems is about 2 seconds with $e$ of about $5* 10^{-9}$.

XDP was run on Jooken's hard file tests [JO22]. On these 2977 tests the average $n=794$, the average $k=202$ and the average fractional error $=2.08* 10^{-4}$. This is the average fractional error from the required optimal result stored with each instance's file.
This average fractional error is similar to $e= 1.13 * 10^{-4}$ from the random trials at the rather higher $k=500$, shown in Table 2. This result suggests XDP's performance on an instance does not significantly depend on a random distribution of profits and weights, since the distributions in these hard file test instances are far from random.

\begin{table}[H]
\caption{XDP: average maximum error $e$ for different $n$ with fixed $k=50$}
\centering
\begin{tabular}{lccr}
\hline\hline
$n$ & $e$  \\
\hline
$10^2$ & $2.65 * 10^{-3}$ \\
$10^3$ & $2.22 * 10^{-3}$ \\
$10^4$ & $2.03 * 10^{-3}$ \\
$10^5$ & $2.07 * 10^{-3}$ \\
\hline
\end{tabular}
\end{table}

\begin{table}[H]
\caption{XDP: average maximum error $e$ for different fixed $k$}
\centering
\begin{tabular}{lcc}
\hline\hline
$k$ & $e$  & $e*k$\\

\hline
$5* 10^0$ & $5.67 * 10^{-2}$ & 0.284 \\
$5* 10^1$ & $2.24 * 10^{-3}$ & 0.112 \\
$5* 10^2$ & $1.13 * 10^{-4}$ & 0.055 \\
$5* 10^3$ & $3.75 * 10^{-6}$ & 0.019 \\
$5* 10^4$ & $1.19 * 10^{-7}$ & 0.006 \\
\hline 
\end{tabular}
\end{table}

\begin{table}[H]
\caption{XDP: average maximum error $e$, runtime and $k$ at each $n$}
\centering
\begin{tabular}{lccr}
\hline\hline
$n$ & $e$  & $secs$ & $k$ \\
\hline
$10^1$ & $6.29* 10^{-2}$ & $1.30 * 10^{-5}$ & $5* 10^0$  \\
$10^2$ & $3.64 * 10^{-3}$ & $7.90 * 10^{-5}$ & $6.1* 10^1$ \\
$10^3$ & $1.58 * 10^{-4}$ & $8.79 * 10^{-4}$ & $6.0* 10^2$ \\
$10^4$ & $5.39 * 10^{-6}$ & $1.09 * 10^{-2}$ & $6.1* 10^3$ \\
$10^5$ & $1.89 * 10^{-7}$ & $1.35 * 10^{-1}$ & $6.0* 10^4$ \\
$10^6$ & $4.59 * 10^{-9}$ & $1.94$ & $6.2* 10^5$ \\

\hline
\end{tabular}
\end{table}

\section {Appendixes}

\subsection {Appendix 1: Maximum possible sum of profits}

A modified version of the familiar greedy algorithm computes $Pmax$, the maximum possible sum of profits for an instance.
It also computes $Gefr$, the greedy algorithm's maximum error at first reject.
It's pseudo code is given below below in Algorithm 2. 

It initially sorts the objects high to low by $p_i/w_i$.  It then adds objects in order until it meets one whose addition would exceed the knapsack weight capacity $c$.
At this first rejected object $r$, it computes this problem's maximum possible sum of profits ($Pmax$) and the greedy algorithm's maximum fractional error at first reject ($Gefr$).
Given the sum of profits added so far $S$ and the sum of weights added so far $W$:

$Pmax= S + (c-W) p_r/w_r$

$Gefr= ((c-W) p_r/w_r) / Pmax$ 

From $Pmax$ we can find the maximum fractional error $e$ for any 0/1 knapsack algorithm whose optimal sum of profits is $S$.

$e=(Pmax-S)/Pmax$

\subsubsection {$Pmax$: experimental validation.}

The maximum profit logic for $Pmax$ was validated on random trials and the hard file tests.
Over one million small $n$ random trials were generated with $n=10..30$. The hard file tests had $n=400..1200$.
For each random trial $Pmax$ was compared to the optimal profit computed by exponential logic. For each hard file test instance $Pmax$ was compared
to the optimal profit recorded with the file data. $Pmax$ was greater than or equal to the optimal sum of profits in every case. 

\begin{algorithm}
\caption{Greedy plus extensions}
\begin{algorithmic}[1]
\Procedure{GreedyPlus}{}

\State $r \gets 0$ \Comment{no object has been rejected yet}
\State $S \gets 0$ \Comment{sum of profits of all objects added to the knapsack}
\State $W \gets 0$ \Comment{sum of weights of all objects added to the knapsack}
\State $x[1..n]\gets 0$	\Comment{no objects are in the knapsack}
\\
\For {$ i=1,n$} \Comment {The objects have been sorted by $p_i/w_i$ high to low}
\If {$(W+w_i <= c)$} 
\State $S \gets S+p_i$
\State $W \gets W+w_i$
\State $x[i] \gets 1$ \Comment{object $i$ is in the knapsack}
\Else
\If {$(r = 0)$} 	\Comment {This is the first rejected object}
\State $r \gets i$
\State $Pmax \gets S +(c-W) p_r/w_r$ 	\Comment {maximum possible profit sum}
\State $Gefr \gets ((c-W) p_r/w_r) / Pmax$ 
\EndIf
\EndIf
\EndFor
\\
\If {$(r = 0)$} \Comment {all $n$ objects are in the knapsack}
\State $Gefr \gets 0$ 
\State $Pmax \gets S$ 
\EndIf
\\
\State $e \gets (Pmax - S)/Pmax$ \Comment {maximum fractional error}
\EndProcedure
\end{algorithmic}
\end{algorithm}

\subsection {Appendix 2: $e$ varies with solution subset size $k$}

In random trials greedy was observed to add very few objects after the first rejected object $r$. It added an average of 2.0 for $k=500$ and 3.6 for $k=50,000$. 
This implies $r \approx k$. In these trials, objects up to $n/2$ were observed to average nearly constant profit and have nearly monotonically increasing average weight. After $n/2$, the average profit nearly monotonically decreased and the average weight was nearly constant.
Assuming all objects up to $r$ average the same profit, the average $Gefr=0.5/k$.
This hypothesis was experimentally confirmed as follows.

Fixed $k$ random trials were made with $k=5..50,000$ and $n=10..100,000$. 
The average $Gefr$ for each $k$ was found from 1000 trials for each possible $n$ for that fixed $k$.
For example: $k$ was set to $5$ and 1000 trials were carried out each for $n=10$, $n=100$, $n=1,000$, $n=10,000$ and $n=100,000$.
The average $Gefr$ reported for $k=5$ is the average found from these 5,000 trials.
The average $Gefr$ for fixed $k$ are reported in Table 4.  
Also shown in this table is the ratio $0.5/k$, which is clearly very similar to the average $Gefr$ for each $k$. 

The $e$ of any algorithm closely related to greedy would be expected to have a similar dependancy. When $T=1$ XDP performs identically to greedy. 

\begin{table}[H] 
\caption{Greedy: average $Gefr$ for fixed $k$}
\centering
\begin{tabular}{lcccr}
\hline\hline
$k$& $Gefr$  & $0.5/k$ \\

\hline
$5* 10^0$ & $1.08 * 10^{-1}$ & $1 * 10^{-1}$ \\
$5* 10^1$ & $1.09 * 10^{-2}$ & $1 * 10^{-2}$ \\
$5* 10^2$ & $1.09 * 10^{-3}$ & $1 * 10^{-3}$  \\
$5* 10^3$ & $1.12 * 10^{-4}$ & $1 * 10^{-4}$ \\
$5* 10^4$ & $1.16 * 10^{-5}$ & $1 * 10^{-5}$  \\
\hline 
\end{tabular}
\end{table}


\begin{thebibliography}{200}

\bibitem[BE53]{BE53} R.Bellman, "Bottleneck problems and dynamic programming", Proceedings of
the National Academy of Sciences of the USA, 39(9): 947–951, 1953.

\bibitem[BL15]{BL15} H. Buhrman, B.Loff, L.Torenvliet, "Hardness of approximation for knapsack problems", Theory of Computing Systems, 56(2): 372-393, 2015

\bibitem[CL25]{CL25} L.Chen,. J.Lian, Y.Mao, G.Zhang, "A Nearly Quadratic-Time FPTAS for Knapsack", arXiv:2308.07821v3 [cs.DS]  2025

\bibitem [CL03]{C03} J.M. Calvin, J.Y.T. Leung, "Average-case analysis of a greedy algorithm for the 0/1 knapsack problem",
Operations Research Letters, 31(2): 202-210, 2003

\bibitem[DA56]{DA56} G.B.Dantzig, "Discrete variable extremum problems", Project Rand, RM1832, 1956

\bibitem[HK17]{HK17} M.Holzhauser, S.O. Krumke, "An FPTAS for the Parametric Knapsack Problem", arXiv:1701.07822v2 [cs.DS] 2017

\bibitem[HS74]{HS74} E.Horowitz, S.Sahni: "Computing partitions with Applications to the Knapsack Problem": Journal of the Association for Computing Machinery, Vol 21(2): 277-292, 1974

\bibitem[IK75]{IK75} O.H.Ibarra, C.E.Kim, "Fast Approximation Algorithms for the Knapsack and Sum of Subset Problems", Journal of the Association for Computing Machinery, 22(4): 463-468, 1975,  

\bibitem[JO22]{JO22} J.Jooken, P.Leyman, and P.De Causmaecker, "A new class of hard problem instances for the 0–1 knapsack problem", European Journal of Operational Research, 301(3): 841–854, 2022

\bibitem[KA72]{KA72} R.M. Karp. "Reducibility among combinatorial problems", Complexity of Computer Computations, The IBM Research Symposia Series: 85–103, 1972

\bibitem[KA23]{KA23} M.Keegan, M.Abolghasemi, "Approximating Solutions to the Knapsack Problem using the Lagrangian Dual Framework", arXiv:2312.03413v1 (cs.LG), 2023

\bibitem[KK04]{KK04} R. Kohli, R.Krishnamurti, P. Mirchandani, "Average performance of greedy heuristics for the integer knapsack problem": European Journal of Operational Research, 154(1): 36-45, 2004

\bibitem[KS14]{KS14} R. E. Korf, E.L. Schreiber, M.D. Moffitt: "Optimal Sequential Multi-Way Number Partitioning", International Symposium on Artificial Intelligence and Mathematics, 2014

\bibitem[MM22]{MM22} F.A. Morales, J.A. Martınez "Expected Performance and Worst Case Scenario Analysis of the Divide-and-Conquer Method for the 0-1 Knapsack Problem": arXiv:2008.04124v2 [cs.DS] 2022

\bibitem[MP99]{MP99} S.Martello, D.Pisinger, P.Toth, "Dynamic programming and strong bounds for the 0-1 knapsack problem", Management Science, 45: 414–24, 1999

\bibitem[PI99]{PI99} D.Pisinger, "Core problems in knapsack algorithms": Operations Research, 47(4): 570-575, 1999.

\bibitem[SS75]{SS75} S.Sahni, "Approximate Algorithms for the 0/1 Knapsack Problem": Journal of the Association for Computing Machinery,  22(1):  115-124, 1975

\bibitem[SS81]{SS81}  R.Schroeppel, A.Shamir: "A T = O(2n/2), S = O(2n/4) algorithm for certain NP-complete problems". SIAM Journal on Computing, 10 (3): 456–464, 1981.


\end{thebibliography}
\end{document}